\documentclass[a4paper,twocolumn]{article}
\usepackage{graphicx}
\usepackage{amsmath}

\begin{document}

\title{Genuine Entanglement of Four Qubit Cluster Diagonal States}
\author{Xiao-yu Chen$^1$, Ping Yu$^2$, Li-zhen Jiang$^1$, Mingzhen Tian$^3$ \\
{\small {$^1$ College of Information and Electronic Engineering, Zhejiang
Gongshang University, Hangzhou, Zhejiang 310018, China }}\\
{\small {$^2$Department of Physics and Astronomy, University of Missouri,
Columbia, Missouri 65211, USA }}\\
{\small {$^3$School of Physics, Astronomy and Computational Sciences, George
Mason University, Fairfax, Virginia 22030, USA}}}
\date{}
\maketitle

\begin{abstract}
We reduce the necessary and sufficient biseparable conditions of the four
qubit cluster diagonal state to concise forms. Only 4 out of the 15
parameters are proved to be relevant in specifying the genuine entanglement
of the state. Using the relative entropy of entanglement as the entanglement
measure, we analytically find the genuine entanglement of all the four qubit
cluster diagonal states. The formulas of the genuine entanglement are of
five kinds, for seven different parameter regions of entanglement.
\end{abstract}

\section{Introduction}

In quantum information and quantum many-body physics, multipartite
entanglement is still a phenomenon that is poorly understood \cite{Horodecki}
\cite{GuhneToth}. Experimentally, multipartite entanglement has been
observed in ion traps \cite{Blatt}, photon polarization \cite{Pan},
superconducting phase or circuit qubit systems \cite{Martinis}, and
nitrogen-vacancy centers in diamond \cite{Neumann}. More than ten qubits
have been entangled in Greenberger-Honre-Zeilinger (GHZ) state in ion traps
\cite{Blatt}, and six-photon cluster state has been observed \cite{Pan1}.
Usually the multipartite entangled states observed in the experiments are
graph states. Graph states are based on graphs and are very useful in
constructing quantum error-correcting codes. Due to noise and imperfections
in preparation, the states prepared are usually mixed states. Typically they
are the so called graph diagonal states. Theoretical research has been
concentrated on the separability and biseparability of the prepared states
\cite{Dur} \cite{GuhneNJP} \cite{GuhneJM} \cite{Doherty} \cite{Eltschka}
\cite{Kay}. Cluster states are special graph states. Recently, necessary and
sufficient biseparable criterion for four qubit cluster diagonal states was
obtained \cite{Guhne}. Biseparable states are multipartite quantum states
that can be expressed as a convex sum of the projectors of product vectors
and bipartite entangled vectors \cite{Acin}. Hence a genuine multipartite
entangled state is not biseparable. Full separable states are those that can
be expressed as a convex sum of the projectors of product vectors. So full
separable state set is the subset of biseparable state set. The criterion
for full separability of a four qubit cluster state is not known. Thus the
quantification of the entanglement is not available when the entanglement
measure involves with the full separable state set.

Quantifying multipartite entanglement is a difficult problem even for a pure
multipartite state. Measures such as entanglement cost, distillable
entanglement work well and have clear operational meanings in bipartite
systems. However, it is not easy to extend them to multipartite systems. The
relative entropy of entanglement (REE) is a valid measure for multipartite
as well as for bipartite systems\cite{Vedral}. For a given quantum state $%
\sigma ,$REE is defined as $E=\min_{\rho \in Sep}S(\sigma \left\|
\rho \right. )$, where $Sep$ is the separable state set, $S(\sigma
\left\| \rho \right. )=Tr(\sigma \log _2\sigma -\sigma \log _2\rho
)$ is the relative entropy. In multipartite system, genuine
entanglement measured by relative entropy can be defined by
minimizing the relative entropy over all biseparable states.
Vedral \textit{et al. }\cite{Vedral} had proposed such a
definition of entanglement for three-partite system. In this
paper, based on the biseparable criterion, we study the genuine
entanglement of four qubit cluster diagonal states measured by
REE.

\section{Cluster diagonal state and its necessary and sufficient biseparable
criteria}

A simple graph $G=(V,\Gamma )$ is composed of a set $V$ of $n$ vertices and
a set of edges characterized by the adjacency matrix $\Gamma .$ The $n\times
n$ symmetric matrix $\Gamma $ has nullified diagonal elements and $\Gamma
_{ij}=$ $1$ if vertices $i$ and $j$ are connected and $\Gamma _{ij}=$ $0$
otherwise. The neighbourhood of a vertex $i$ is denoted by $N_i=\{j\in
V\left| \Gamma _{ij}=1\right. $ $\}.$ Consider a system of $n$ qubits, and
define the mutually commutating stabilizer operators:
\begin{equation}
K_i=X_i\prod_{j\in N_i}Z_j  \label{rtt1}
\end{equation}
where $X_i$ and $Z_j$ are Pauli $X$ and $Z$ matrices at vertices $i$ and $j,$
respectively. The operators stabilize the graph state
\begin{equation}
\left| G\right\rangle =\frac 1{\sqrt{2^n}}\sum_{\mu =\mathbf{0}}^{\mathbf{1}%
}(-1)^{\frac 12\mu \Gamma \mu ^T}\left| \mu \right\rangle  \label{rtt2}
\end{equation}
such that $K_i\left| G\right\rangle =\left| G\right\rangle .$ Graph state
basis are $\left| G_{\alpha _1\alpha _2\ldots \alpha _n}\right\rangle
=Z_1^{\alpha _1}Z_2^{\alpha _2}\cdots Z_n^{\alpha _n}\left| G\right\rangle $
with $\alpha _i=0,1,$ each of them is the common eigenstate of all the
operators $K_i,$ with eigenvalues $\pm 1.$

The graph of the four qubit cluster state $\left| Cl\right\rangle $ has
three edges such that $\Gamma _{i,i+1}=\Gamma _{i+1,i}=$ $1$ $(i=1,2,3)$ and
all the other elements of $\Gamma $ are $0$. The stabilizer operators are
\begin{eqnarray}
K_1 &=&X_1Z_2I_3I_4,\text{ }K_2=Z_1X_2Z_3I_4,  \label{rtt3} \\
K_3 &=&I_1Z_2X_3Z_4,\text{ }K_4=I_1I_2Z_3X_4.  \label{rtt4}
\end{eqnarray}
where $I_i$ is the identity matrix for vertex $i$. The four qubit cluster
basis states are $\left| Cl_{\alpha _1\alpha _2\alpha _3\alpha
_4}\right\rangle $ with $K_i\left| Cl_{\alpha _1\alpha _2\alpha _3\alpha
_4}\right\rangle =(-1)^{\alpha _i}\left| Cl_{\alpha _1\alpha _2\alpha
_3\alpha _4}\right\rangle .$ A four qubit cluster diagonal state $\sigma $
is the probability mixture of the cluster basis states
\begin{equation}
\sigma =\sum_{\alpha ,\beta ,\gamma ,\delta =0}^1F_{\alpha \beta \gamma
\delta }\left| Cl_{\alpha \beta \gamma \delta }\right\rangle \left\langle
Cl_{\alpha \beta \gamma \delta }\right| ,  \label{rtt5}
\end{equation}
where $F_{\alpha \beta \gamma \delta }\geq 0$ and $\sum_{\alpha ,\beta
,\gamma ,\delta =0}^1F_{\alpha \beta \gamma \delta }=1.$

The original necessary and sufficient biseparable criteria are \cite{Guhne}
\begin{eqnarray}
2F_{\alpha \beta \gamma \delta } &\leq &\sum_{\xi ,\eta =0}^1(F_{\alpha \xi
\eta \delta }+F_{\alpha \xi \eta \overline{\delta }}+F_{\overline{\alpha }%
\xi \eta \delta })  \label{wee1} \\
2F_{\alpha \beta \gamma \delta }+2F_{\overline{\alpha }\mu \nu \overline{%
\delta }} &\leq &\sum_{\xi ,\eta =0}^1(F_{\alpha \xi \eta \delta }+F_{\alpha
\xi \eta \overline{\delta }}  \nonumber \\
&&+F_{\overline{\alpha }\xi \eta \delta }+F_{\overline{\alpha }\xi \eta
\overline{\delta }})  \label{wee2}
\end{eqnarray}
for all the subscripts $\alpha ,\beta ,\gamma ,\delta ,\mu ,\nu =0,1$, where
$F_{\alpha \beta \gamma \delta }=\left\langle C_{4\alpha \beta \gamma \delta
}\right| \sigma \left| C_{4\alpha \beta \gamma \delta }\right\rangle .$
Violation of the biseparable criteria means genuine entanglement. Note that
the right hand side of inequality (\ref{wee2}) is just equal to 1 according
to the normalization of $\sigma $ in cluster state basis. Hence the two
criteria can be written as
\begin{eqnarray}
2F_{\alpha \beta \gamma \delta }+\sum_{\xi ,\eta =0}^1F_{\overline{\alpha }%
\xi \eta \overline{\delta }} &\leq &1,  \label{wee3} \\
F_{\alpha \beta \gamma \delta }+F_{\overline{\alpha }\mu \nu \overline{%
\delta }} &\leq &\frac 12,  \label{wee4}
\end{eqnarray}
Denote
\begin{eqnarray}
p_{2\alpha +\delta } &=&\max_{\beta ,\gamma }\{F_{\alpha \beta \gamma \delta
}\},  \label{rtt6} \\
p_{4+2\alpha +\delta } &=&\sum_{\beta ,\gamma }F_{\alpha \beta \gamma \delta
}-p_{2\alpha +\delta }.  \label{rtt7}
\end{eqnarray}
Then inequalities (\ref{wee3}) and (\ref{wee4}) can be further reduced to
\begin{eqnarray}
p_0+p_3 &\leq &\frac 12,  \label{wee5} \\
2p_0+p_3+p_7 &\leq &1,  \label{wee6} \\
2p_3+p_0+p_4 &\leq &1,\text{ }  \label{wee7} \\
p_1+p_2 &\leq &\frac 12,\text{ }  \label{wee8} \\
2p_1+p_2+p_6 &\leq &1,\text{ }  \label{wee9} \\
2p_2+p_1+p_5 &\leq &1,\text{ }  \label{wee10}
\end{eqnarray}
We will prove that inequality (\ref{wee4}) and inequalities (\ref{wee5}), (%
\ref{wee8}) are equivalent. For any $\beta ,\gamma ,\mu ,\nu ,$ if (\ref
{wee4}) is violated, then one of (\ref{wee5}) and (\ref{wee8}) should be
violated from the definition of $p_{2\alpha +\delta }$. If (\ref{wee5})
and/or (\ref{wee8}) are violated, then at least one case of (\ref{wee4}) is
violated for some $\beta ,\gamma ,\mu ,\nu ,$ since (\ref{wee5}) and (\ref
{wee8}) are special cases of (\ref{wee4}). The same reasoning can be applied
to prove the equivalence of inequality (\ref{wee3}) and inequalities (\ref
{wee6})(\ref{wee7})(\ref{wee9})(\ref{wee10}).

If inequality (\ref{wee5}) is violated, then inequalities (\ref{wee8})-(\ref
{wee10}) are preserved due to the normalization $\sum_{\alpha ,\beta ,\gamma
,\delta }F_{\alpha \beta \gamma \delta }=\sum_{\alpha ,\delta }(p_{2\alpha
+\delta }+p_{4+2\alpha +\delta })=1.$ If inequality (\ref{wee6}) is
violated, we also have inequalities (\ref{wee8})-(\ref{wee10}) been
fulfilled. We conclude that if one of the inequalities (\ref{wee5})-(\ref
{wee7}) is violated, then inequalities (\ref{wee8})-(\ref{wee10}) are all
preserved and vice versa. So we only need to consider half of the
inequalities been violated. We consider the violation of inequalities (\ref
{wee5})-(\ref{wee7}), in the following we will mainly work on the parameters
$p_0,p_3,p_4,p_7.$ Suppose, for example, the maximal of $F_{0\beta \gamma 0}
$ ($\beta ,\gamma =0,1$) be $F_{0000},$ the maximal of $F_{1\beta \gamma 1}$
($\beta ,\gamma =0,1$) be $F_{1001}$, then $p_0=F_{0000}$ , $p_3=$ $%
F_{1001}, $ $p_4=F_{0010}+F_{0100}+F_{0110},$ $p_7=$ $%
F_{1011}+F_{1101}+F_{1111}.$

\section{Entanglement measure}

The (genuine) REE \cite{Vedral} defined for a tripartite entangled
state can easily be extended to a generic multipartite state. The
genuine entanglement of a genuine entangled state $\sigma $
measured by the REE is
\begin{eqnarray}
E &=&\min_{\rho \in Bisep}S(\sigma \left\| \rho \right. )  \nonumber \\
&=&Tr(\sigma \log _2\sigma -\sigma \log _2\varrho ),  \label{wee11}
\end{eqnarray}
where $Bisep$ is the biseparable set, $\varrho $ is the closest
biseparable state, namely, biseparable state that minimizes the
relative entropy. The genuine REE measures how `far' is the
genuine entangled state from its nearest biseparalbe state, i.e,
the state that is not genuine entangled. For a cluster diagonal
state $\sigma $, it is easy to show that $\varrho $ is also a
cluster diagonal state following the reasoning of declaration that
the closest separable state of a Bell diagonal entangled state is
a Bell diagonal state \cite{Vedral}. Let $\varrho =\sum_{\alpha
,\beta ,\gamma ,\delta =0}^1\Lambda _{\alpha \beta \gamma \delta
}\left| Cl_{\alpha \beta \gamma \delta }\right\rangle \left\langle
Cl_{\alpha \beta \gamma \delta }\right| $, with $\left\{ \Lambda
_{\alpha \beta \gamma \delta }\right\} $ forming a probability
distribution. The genuine REE is
\begin{equation}
E=\sum_{\alpha ,\beta ,\gamma ,\delta =0}^1F_{\alpha \beta \gamma \delta
}\log _2\frac{F_{\alpha \beta \gamma \delta }}{\Lambda _{\alpha \beta \gamma
\delta }}.  \label{wee12}
\end{equation}
The convex property of $-\log $ function means that the closest biseparable
state $\varrho $ should be at the boundary of the biseparable state set.
Denote $\lambda _{2\alpha +\delta }=\max_{\beta ,\gamma }\{\Lambda _{\alpha
\beta \gamma \delta }\},\lambda _{4+2\alpha +\delta }=\sum_{\beta ,\gamma
}\Lambda _{\alpha \beta \gamma \delta }-\lambda _{2\alpha +\delta },$ then
at least one of the equality should be reached in the biseparable criteria
of the state $\varrho ,$ namely, if we replace $p_i$ with $\lambda _{i\text{
}}$for inequalities (\ref{wee5})-(\ref{wee10}), then at least one of the
inequalities should be an equation. We use $p_i$ to specify the possibly
genuine entangled state $\sigma ,$ use $\lambda _{i\text{ }}$to specify the
closest biseparable state $\varrho .$

\begin{figure}[tbp]
\includegraphics[ trim=0.000000in 0.000000in -0.138042in 0.000000in,
height=2.5in, width=3.5in ]{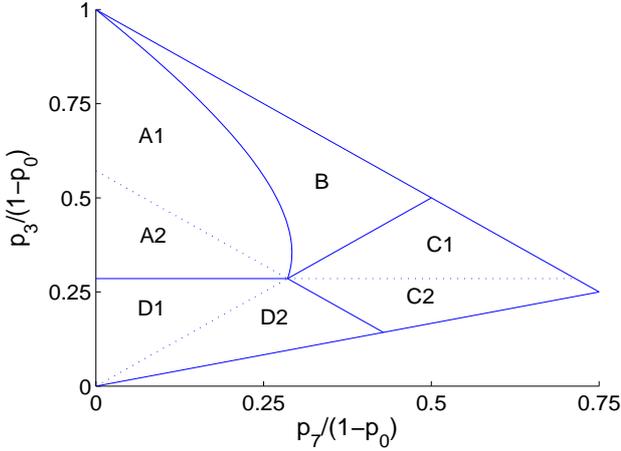}
\caption{(Color online) Edge lines or curve and genuine entanglement or
biseparable regions for $p_0=0.3$. Genuine entangled state Regions: $A_{1}$
violates $(\ref{wee5})$ and $(\ref{wee6})$, $A_{2}$ violates $(\ref{wee5})$,
the genuine REE can be $E_{A},E_{A^{\prime\prime}}$ or $%
E_{A^{\prime\prime\prime}}$ due to the fourth parameter $p_{4}$;
$B$
violates $(\ref{wee5})$ and $(\ref{wee6})$, the genuine REE is $%
E_{B}$; $C_{1}$ violates $(\ref{wee5})$ and $(\ref{wee6})$,
$C_{2}$ violates $(\ref{wee6})$, the genuine REE is $E_{C}$.
Biseparable state region: $D_{2}$. Region whose biseparablity yet
to be determined by the fourth parameter: $D_{1}$. In regions
$D_{1}$,$D_{2}$, both $(\ref{wee5}) $ and $(\ref{wee6})$ are
preserved. For all the regions, possible violation of
$(\ref{wee7})$ will be discussed in the next section.}
\end{figure}

\begin{figure}[tbp]
\includegraphics[ trim=0.000000in 0.000000in -0.138042in 0.000000in,
height=2.5in, width=3.5in ]{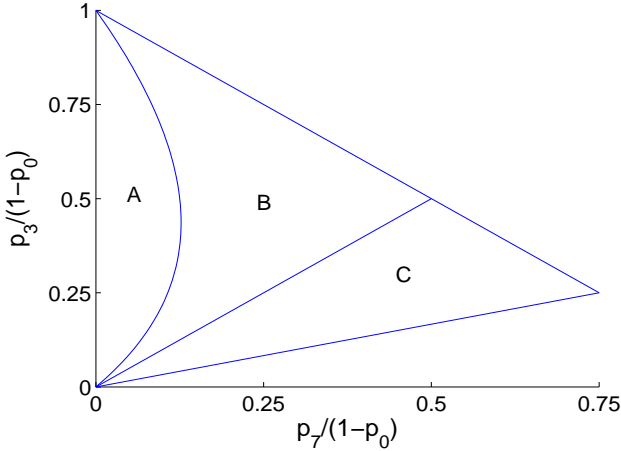} \caption{(Color
online)Edge lines or curve of genuine entanglement regions for
$p_0=0.6$. Genuine entangled state Regions: $A$,$B$,$C$, both
$(\ref {wee5})$ and $(\ref{wee6})$ are violated. The genuine REE
for
region $A$ can be $E_{A},E_{A^{\prime\prime}}$ or $%
E_{A^{\prime\prime\prime}}$ due to the fourth parameter $p_{4}$,
The genuine REE is $E_{B}$ for $B$, is $E_{C}$ for $C$. }
\end{figure}

\section{Three parameter states}

Suppose a genuine entangled state $\sigma $ violate inequality (\ref{wee5})
or inequality (\ref{wee6}) or both of them, while leaving inequality (\ref
{wee7}) pending for further consideration in the next section. Hence $\sigma
$ is characterized by three parameters $p_0,p_3,p_7$ concerning its genuine
entanglement property. Without considering the fourth parameter $p_4$ or
possible violation of inequality (\ref{wee7}), the genuine entanglement
obtained in this section is only the candidate of the final result.

When inequality (\ref{wee5}) is violated, namely, $p_0+p_3>\frac
12$,
the parameter regions can be shown as $A_1,A_2,B$ and $C_1$ in Fig.1 for $%
p_0<\frac 12$ or $A,B,C$ in Fig.2 for $p_0\geq \frac 12.$ When
inequality (\ref{wee6}) is violated, namely, $2p_0+p_3+p_7>1,$ the
parameter regions can be shown as $A_2,B,$ $C_1$ and $C_2$ in
Fig.1 for $p_0<\frac 12$ or $A,B,C$ in Fig.2 for $p_0\geq \frac
12.$

We first consider the case of $p_0<\frac 12.$ For any give genuine entangled
state in regions $A_1,A_2,B,C_1$ and $C_2,$ there are three classes of
possible closest biseparable states.

Class I is the closest biseparable state set with
\begin{equation}
\lambda _0+\lambda _3=\frac 12.  \label{wee13}
\end{equation}
Using Lagrange multiplier to minimize the relative entropy, we can determine
\begin{eqnarray}
\lambda _0 &=&\frac{p_0}{2(p_0+p_3)},\text{ }\lambda _3=\frac{p_3}{2(p_0+p_3)%
}  \label{wee14} \\
\Lambda _{\alpha \beta \gamma \delta } &=&\frac{F_{\alpha \beta \gamma
\delta }}{2(1-p_0-p_3)},\text{ for all the others }  \label{wee15}
\end{eqnarray}
and the genuine REE is
\begin{equation}
E_A=1-H_2(p_0+p_3).  \label{wee16}
\end{equation}
where $H_2(x)=-x\log _2x-(1-x)\log _2(1-x)$ is the binary entropy function.

Class II is the closest biseparable state set with
\begin{equation}
2\lambda _0+\lambda _3+\lambda _7=1.  \label{wee18}
\end{equation}
We obtain the solution
\begin{eqnarray}
\lambda _0 &=&\frac 12(1-p_3-p_7),\text{ }\lambda _3=p_3,\text{ }\lambda
_7=p_7,  \label{wee19} \\
\Lambda _{\alpha \beta \gamma \delta } &=&\frac{(1-p_3-p_7)F_{\alpha \beta
\gamma \delta }}{2(1-p_0-p_3-p_7)},\text{ for all the others}  \label{wee20}
\end{eqnarray}
and the genuine REE
\begin{equation}
E_C=(1-p_3-p_7)[1-H_2(\frac{p_0}{1-p_3-p_7})].  \label{wee21}
\end{equation}

Class III is the closest biseparable state set with both (\ref{wee13}) and (%
\ref{wee18}). The Lagrange multiplier solution of minimizing
(\ref{wee12}) with restrictions of (\ref{wee13}) and (\ref{wee18})
and the normalization condition is
\begin{eqnarray}
\lambda _0 &=&\frac 12(1-p_3-p_7),\text{ }\lambda _3=\text{ }\lambda
_7=\frac 12(p_3+p_7),  \label{wee22} \\
\Lambda _{\alpha \beta \gamma \delta } &=&\frac{(1-p_3-p_7)F_{\alpha \beta
\gamma \delta }}{2(1-p_0-p_3-p_7)},\text{ for all the others}  \label{wee23}
\end{eqnarray}
The genuine REE is
\begin{eqnarray}
E_B &=&1-(p_3+p_7)H_2(\frac{p_3}{p_3+p_7})  \nonumber \\
&&-(1-p_3-p_7)H_2(\frac{p_0}{1-p_3-p_7})].  \label{wee24}
\end{eqnarray}

For class I closest biseparable states, we should check if
$2\lambda _0+\lambda _3+\lambda _7\leq 1$ is preserved. Otherwise,
the supposed closest state is not biseparable, the solution
(\ref{wee14})-(\ref{wee16}) is no longer proper. Using the
solution (\ref{wee14})-(\ref{wee15}), we have $2\lambda
_0+\lambda _3+\lambda _7=\frac 12(1+\frac{p_0}{p_0+p_3}+\frac{p_7}{1-p_0-p_3}%
),$ which is less than or equal to $1$ when
\begin{equation}
p_7\leq \frac{1-p_0-p_3}{p_0+p_3}p_3\equiv p_{AB}.  \label{wee17}
\end{equation}
This characterizes the region $A$ (the union $A_1$ and $A_2$ in
Fig.1). The equality in (\ref{wee17}) represents the border curve
in Fig.1 or Fig.2 that divides region $A$ and region $B.$

For class II closest biseparable states, we should check whether
$\lambda _0+\lambda _3\leq \frac 12$ is preserved. If it is not,
the supposed closest state will
go out of the biseparable state set, the solution (\ref{wee19})-(\ref{wee21}%
) is not valid. Using (\ref{wee22}), we have $\lambda _0+\lambda _3=\frac
12(1+p_3-p_7)$, which is less than or equal to $\frac 12$ only when
\begin{equation}
p_7\geq p_3.  \label{wee25}
\end{equation}
This characterizes region $C$ (the union $C_1$ and $C_2$ in
Fig.1). The equality in (\ref{wee25}) represents the border line
in Fig.1 or Fig.2 that divides region $C$ and region $B.$

All the states in region $A$ violate (\ref{wee25}), thus they have
not closest biseparable states of class II. The possible solutions
are class I and class III with genuine entanglement $E_A$ and
$E_B,$respectively. Since $1-H_2(x)$ is a convex function, we have
$E_B\geq E_A$, the genuine REE of region $A$ is $E_A$.

All the states in region $C$ violate (\ref{wee17}), so they have
not closest biseparable states of class I. The possible solutions
are class II and class III with genuine entanglement $E_C$ and
$E_B,$respectively. However, $E_C$ is only a part of $E_B,$ we
have $E_B\geq E_C$. Hence, the genuine entanglement of region $C$
is $E_C.$

The states in region $B$ (the states at the border curve with $A$ and border
line with $C$ are not included) violate either (\ref{wee17}) or (\ref{wee25}%
). Their closest biseparable states can not be in class I or class
II. They are in class III. The genuine REE of region $B$ is $E_B.$

The case of $p_0\geq \frac 12$ can be similarly analyzed. In Fig.2, all the
states in regions $A,B,C$ are genuinely entangled.

The conclusion for three parameter states is: the candidate
genuine REE in regions $A,B$ and $C$ is $E_A,E_B$ and $E_C.$ Note
that $p_3$ is the maximum of $F_{1\beta \gamma 1}$, so $3p_3\geq
p_7$, Hence the region with $p_3<\frac 13p_7$ is meaningless.

The closest biseparable states of the genuine entangled state in
region $A$ and $C$ are on the lines of (\ref{wee13}) and
(\ref{wee18}) (hyperplanes in fact ), respectively. The closest
biseparable states of the genuine entangled state in region $B$
are on the intersection point of the lines (\ref{wee13}) and (\ref
{wee18}).

\section{Four parameter states}

The regions are classified as $A,B,C,D$ with parameters
$p_0,p_3,p_7$ in last section. We will consider the further
classification of each regions by the fourth parameter $p_4.$ The
biseparable state set is shown in Fig.3 for $\lambda _0=0.2.$ The
three classes of closest biseparable state sets defined in last
section are shown in Fig.3 with surfaces I, II and intersection
line III. There are two new classes of closest biseparable states
appear when the fourth parameter $p_4$ is considered.

Class IV (shown in Fig.3 with surface IV) is the closest
biseparable state set with
\begin{equation}
\lambda _0+2\lambda _3+\lambda _4=1.  \label{att8}
\end{equation}
The Lagrange multiplier solution of minimizing (\ref{wee12}) with
restrictions of (\ref {att8}) is
\begin{eqnarray}
\lambda _3 &=&\frac 12(1-p_0-p_4),\text{ }\lambda _0=p_0,\text{ }\lambda
_4=p_4,  \label{att9} \\
\Lambda _{\alpha \beta \gamma \delta } &=&\frac{(1-p_0-p_4)F_{\alpha \beta
\gamma \delta }}{2(1-p_0-p_3-p_4)},\text{ for all the others}  \label{att10}
\end{eqnarray}
The genuine REE is
\begin{equation}
E_{A^{\prime \prime \prime }}=(1-p_0-p_4)[1-H_2(\frac{p_3}{1-p_0-p_4})].
\label{wee26}
\end{equation}

Class V (shown in Fig.3 with line V) is the closest biseparable
state set with both (\ref {wee13}) and (\ref{att8}). The solution
of (\ref{wee12}) is
\begin{eqnarray}
\lambda _3 &=&\frac 12(1-p_0-p_4),\text{ }\lambda _0=\text{ }\lambda
_4=\frac 12(p_0+p_4),  \label{att11} \\
\Lambda _{\alpha \beta \gamma \delta } &=&\frac{(1-p_0-p_4)F_{\alpha \beta
\gamma \delta }}{2(1-p_0-p_3-p_4)},\text{ for all the others}  \label{att12}
\end{eqnarray}
The genuine REE is
\begin{eqnarray}
E_{A^{\prime \prime }} &=&1-(p_0+p_4)H_2(\frac{p_0}{p_0+p_4})  \nonumber \\
&&-(1-p_0-p_4)H_2(\frac{p_3}{1-p_0-p_4})].  \label{wee28}
\end{eqnarray}

The genuine entanglement regions in the four parameter system with given $%
p_0 $ are determined by parameters $\frac{p_3}{1-p_0},\frac{p_7}{1-p_0},%
\frac{p_4}{1-p_0}.$ All the genuine entanglement regions described in Fig.1
and Fig.2 now are three dimensional when $p_4$ is considered. The regions
are three dimensional by adding the third dimension $p_4$ to the two
dimensional graphs shown in Fig.1 or Fig.2. For example, the bottom of the
three dimensional region $A$ is the graph in Fig.1 or Fig.2. The roof of $A$
is determined by the condition
\begin{equation}
p_0+p_4+p_3+p_7=1,  \label{att13}
\end{equation}
which comes from the normalization condition $1=\sum_{\alpha ,\beta ,\gamma
,\delta }F_{\alpha \beta \gamma \delta }\geq p_0+p_4+p_3+p_7$. The other
border surfaces are $p_7=0,$ $p_7=p_{AB}$ and $p_0+p_3=\frac 12.$

For region $A,$ we have proven at last section that the closest
biseparable states can not be in class II. Suppose the closest
biseparable states are in class I, the solution is
(\ref{wee14})-(\ref{wee16}). Then $\lambda _0+2\lambda _3+\lambda
_4=\frac 12(1+\frac{p_3}{p_0+p_3}+\frac{p_4}{1-p_0-p_3}).$ The
biseparable condition $\lambda _0+2\lambda _3+\lambda _4\leq 1$ is
preserved only when
\begin{equation}
p_4\leq \frac{1-p_0-p_3}{p_0+p_3}p_0\equiv p_{A^{\prime }A^{\prime \prime }}.
\label{wee31}
\end{equation}
We denote the region in $A$ limited by (\ref{wee31}) as $A^{\prime }$.

For the other part of $A$, suppose the closest biseparable states
be in class IV, we have the solution (\ref{att9})-(\ref{wee26}).
Then $\lambda _0+\lambda _3=\frac 12(1+p_0-p_4)$, which is less
than or equals to $\frac 12$ when
\begin{equation}
p_4\geq p_0.  \label{att14}
\end{equation}
We denote the region in $A$ limited by (\ref{att14}) as $A^{\prime \prime
\prime }$. In region $A,$ we have $p_0+p_3>\frac 12,$ from which we can
derive $p_0>p_{A^{\prime }A^{\prime \prime }}.$ Hence $A^{\prime }$ and $%
A^{\prime \prime \prime }$ do not overlap. The region in $A$ with $%
p_0<p_4<p_{A^{\prime }A^{\prime \prime }}$ then is denoted as $A^{\prime
\prime }$. Hence the region $A$ is divided into three layers from bottom to
top as $A^{\prime },A^{\prime \prime }$ and $A^{\prime \prime \prime }$ along $%
p_4$ direction. The layers are shown in Fig.4 without considering
parameter $p_7$.

For layer $A^{\prime \prime },$ the closest biseparable states
neither belong to class I nor class IV. We consider class V. The
solution then is (\ref{att11})-(\ref {wee28}). It is possible that
$A^{\prime }$ and $A^{\prime \prime \prime }$ may have class V
closest biseparable states. However, we have $E_A\leq E_{A^{\prime
\prime }}$ due to the convexity of the function $1-H_2(x).$ We have $%
E_{A^{\prime \prime \prime }}\leq E_{A^{\prime \prime }}$ for
$E_{A^{\prime \prime \prime }}\ $ is a part of $E_{A^{\prime
\prime }}.$ Thus closest biseparable states of $A^{\prime }$ and
$A^{\prime \prime \prime }$ can not be in class V.

We should further check if the closest biseparable states of $A$
are in class III. We have proven that $E_A\leq E_B$ in last
section, the proof is valid for layer $A^{\prime }.$ We only need
to consider $A^{\prime \prime }$ and $A^{\prime
\prime \prime }.$ The closest biseparable states for layers $A^{\prime \prime }$ and $%
A^{\prime \prime \prime }$ can not be class III (see appendix).
The genuine REE is $E_A,E_{A^{\prime \prime }},E_{A^{\prime \prime
\prime }}$ for $A^{\prime },A^{\prime \prime },A^{\prime \prime
\prime },$respectively.

For all states in region $A,$ inequality (\ref{wee5}) is violated. In layer $%
A^{\prime },$ inequalities (\ref{wee6}) and (\ref{wee7}) can be violated or
not due to the location of the state. In layer $A^{\prime \prime }$, (\ref
{wee6}) may or may not be violated due to the location of the state. In
layer $A^{\prime \prime \prime },$ (\ref{wee6}) is preserved since $1\geq
p_0+p_4+p_3+p_7$ and $p_4\geq p_0.$ In both layers $A^{\prime \prime }$ and $%
A^{\prime \prime \prime },$ (\ref{wee7}) is violated as shown in Fig.4.

Consider the closest biseparable states of class IV, we have
$p_4\geq p_0$ in order to preserve $\lambda _0+\lambda _3\leq
\frac 12.$ When (\ref{wee6}) is violated, as in regions $B$ and
$C,$ we have $2p_0+p_3+p_7>1\geq p_0+p_4+p_3+p_7,$ the last
inequality comes from the normalization condition. Thus $p_0>p_4.$
So regions $B$ and $C$ have not class IV solution. Class V is also
not a solution for them (see appendix). The genuine REE keeps the
same form as in the three parameter situation
of last section for regions $B$ and $C$. In regions $B$ and $C,$ inequality (%
\ref{wee6}) is violated, in regions $B$ and $C_1,$inequality (\ref{wee5}) is
violated. Inequality (\ref{wee7}) can either be violated or preserved in
region $B.$ In regions $C$ and $D_2$, the inequality (\ref{wee7}%
) is preserved. Note that $p_0+p_4+p_3+p_7\leq 1$, we have $p_0+2p_3+p_4\leq
1+p_3-p_7,$ thus if $p_3\leq p_7,$ we should have inequality (\ref{wee7}).
Region $D_2$ is biseparable even considering the fourth parameter $p_4.$

Violation of inequality (\ref{wee7}) is possible for regions $D_1$ as shown
in Fig.4. Region $D_1$ is divided into two layers $D_1^{\prime }$ and $%
D_1^{\prime \prime },$ with $D_1^{\prime \prime }$ biseparable and $%
D_1^{\prime }$ genuine entangled. Consider the class VI\ closest
biseparable states for
layer $D_1^{\prime }$, the genuine entanglement of layer $D_1^{\prime }$ is $%
E_{A^{\prime \prime \prime }}.$ Closest biseparable states in
class I can not be the solution since $D_1^{\prime }$ does not
overlap with $A^{\prime }.$ Closest biseparable states in class V
can not be the solution since $E_{A^{\prime \prime \prime }}\leq
E_{A^{\prime \prime }}$. Class II solution requires $p_7\geq p_3$,
while in $D_1^{\prime }$ we have $p_7<p_3.$ Class III is also not
the solution for $D_1^{\prime }$ (see appendix).

\begin{figure}[tbp]
\includegraphics[ trim=0.000000in 0.000000in -0.138042in 0.000000in,
height=2.5in, width=3.5in ]{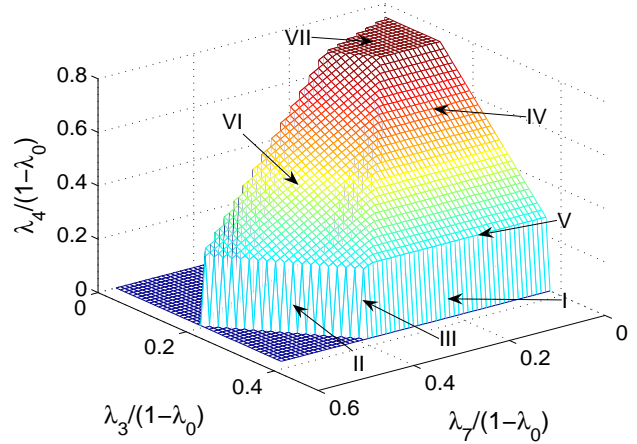}
\caption{(Color online) Border surfaces of biseparable state set for $%
\lambda_0=0.2$. Surfaces I,II,IV are for $\lambda_{0}+\lambda_{3}=1/2$, $%
2\lambda_{0}+\lambda_{3}+\lambda_{7}=1$, $\lambda_{0}+2\lambda_{3}+%
\lambda_{4}=1$, respectively, they are the border surfaces of biseparable
and genuine entangled states. Surfaces VI and VII are for $%
\lambda_{0}+\lambda_{3}+\lambda_{4}+\lambda_{7}=1$ and $\lambda_{4}=3%
\lambda_{0}$, they and the surface $\lambda_{7}=3\lambda_{3}$ are the border
surfaces of biseparable states and the unphysical region. Lines III and V
are the intersections of the surfaces. }
\end{figure}

\begin{figure}[tbp]
\includegraphics[ trim=0.000000in 0.000000in -0.138042in 0.000000in,
height=2.5in, width=3.5in ]{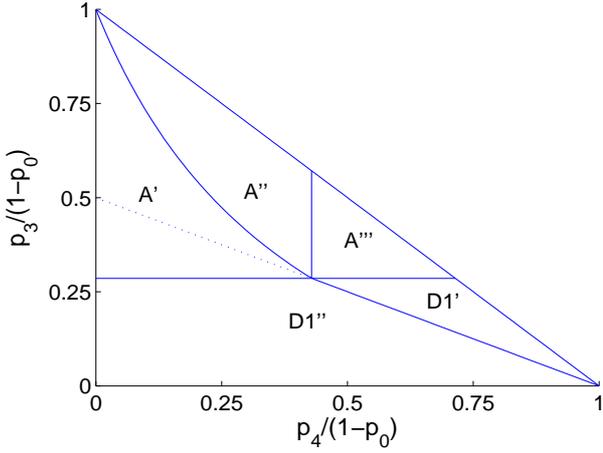}
\caption{(Color online)Edge lines or curve of genuine entanglement regions
for $p_0=0.3$. Genuine entanglement regions $A^{\prime },A^{\prime \prime }$
are based on the regions $A_1$ or $A_2$, region $A^{\prime \prime \prime }$
is based on $A_2$ in Fig.1, further characterized by the fourth parameter $%
p_4$. The entanglement is $E_{A},E_{A^{\prime \prime }},E_{A^{\prime \prime
\prime }},$ respectively. Biseparable region $D_1^{\prime \prime }$ and
genuine entanglement region $D_1^{\prime }$ are based on region $D_1$ in
Fig.1, further characterized by $p_4$. The dash line is the border of
inequality (\ref{wee7}). }
\end{figure}

The situation of $p_4>3p_0$ does not exist due to the assuming of $%
p_0=\max_{\beta ,\gamma }F_{0\beta \gamma 0}.$

The parameter regions and the corresponding genuine REE are
summarized in Table 1. All the regions should be subjected to the
normalization constrain $p_4+p_0+p_3+p_7\leq 1$ and $p_4\leq
3p_0,$ $p_7\leq 3p_3.$

\begin{table}[tbp]
\begin{center}
{\bfseries The genuine entangled or biseparable regions}\\[1ex]
\par
\begin{tabular}{|l||l|l|l|}
\hline
\text{} & \text{Descriptions} & \text{ E} & V \\ \hline
$A^{\prime }$ & $p_4\leq p_{A^{\prime }A^{\prime \prime }}; p_7\leq p_{AB}; $
& $E_A$ & $(\ref{wee5})(\ref{wee6})^{*}$ \\
\text{ } & $\text{ }p_3>\frac 12-p_0$ &  & $(\ref{wee7})^{*}$ \\ \hline
$A^{\prime \prime }$ & $p_{A^{\prime }A^{\prime \prime }}<p_4 < p_0;\text{ }%
\text{ } p_7\leq p_{AB};$ & $E_{A^{\prime\prime }}$ & $(\ref{wee5})(\ref
{wee6})^{*}$ \\
& $\text{ }p_3>\frac 12-p_0$ &  & (\ref{wee7}) \\ \hline
$A^{\prime \prime \prime }$ & $p_4\geq p_0;\text{ }\text{ } p_7\leq p_{AB};$
& $E_{A^{\prime\prime \prime}}$ & (\ref{wee5}) \\
& $\text{ }p_3>\frac 12-p_0$ &  & (\ref{wee7}) \\ \hline
B & $p_3> p_7>p_{AB}$;\text{ }$p_3>\frac 12-p_0 $ & $E_B$ & (\ref{wee5})(\ref
{wee6}) \\
&  &  & $(\ref{wee7})^{*}$ \\ \hline
$C_1$ & $p_7\geq p_3$;\text{ }$p_3>\frac 12-p_0 $ & $E_C $ & (\ref{wee5})(%
\ref{wee6}) \\ \hline
$C_2$ & $p_7>1-2p_0-p_3$; & $E_C $ & (\ref{wee6}) \\
& $p_3\leq \frac 12-p_0 $ &  &  \\ \hline
$D_1^{\prime }$ & $p_4>1-p_0-2p_3;\text{ }p_7<p_3$; & $E_{A^{\prime\prime
\prime}}$ & (\ref{wee7}) \\
& $p_3\leq \frac 12-p_0$ &  &  \\ \hline $D_1^{\prime \prime }$ &
$p_4\leq 1-p_0-2p_3$;\text{ }$p_7<p_3$; & $0^{*}$ & None
\\
& $p_3\leq \frac 12-p_0$ &  &  \\ \hline
$D_2$ & $1-2p_0-p_3\geq p_7\geq p_3$; & $0^{*}$ & None \\
& $p_3\leq \frac 12-p_0$ &  &  \\ \hline
\end{tabular}
\\[0.5ex]
\end{center}
\caption{The regions classified by their parameters. Here E is the genuine
entanglement. V means violation of inequalities (\ref{wee5})(\ref{wee6})and (%
\ref{wee7}). $(\ref{wee6})^{*}$ or $(\ref{wee7})^{*}$ represents
that $(\ref {wee6})$ or $(\ref{wee7})$ is violated in some part of
the region and preserved in the other part of the region. $0^{*}$
is 0 only when $(\ref {wee8})$-$(\ref {wee10})$ are preserved. }
\end{table}

\section{Conclusions and Discussions}

We have completely solved the genuine entanglement problem of four
qubit cluster diagonal state. For any probability mixture of four
qubit cluster basis states, we first reduce the total number of
parameters from 15 to 8. Then we have proven that at most half of
the necessary and sufficient biseparable criteria of are violated
for any genuine entangled state, the number of the parameters
involved then is further reduced to 4. The four parameter states
are classified as biseparable and genuine entangled. The
entanglement
measure for a genuine entangled state $\sigma $ is the relative entropy of $%
\sigma $ with respect to the closest biseparable state. We found
that the closest biseparable state set is on the interface of
biseparable state set and genuine entangled state set. The closest
biseparable state set is divided into five classes. We classify
the genuine entangled states into several regions and find their
closest biseparable state classes. Each region has its closest
biseparable state set. The entanglement is given analytically for
any genuine entangled four qubit cluster diagonal state.

The four parameters are symmetry in some sense. For $p_0+p_3>\frac 12,$ the
parameter region is divided into five subregions: $A^{\prime \prime \prime
},A^{\prime \prime },A^{\prime },B$ and $C_1,$ where $C_1$ and $A^{\prime
\prime \prime }$ are symmetric, $B$ and $A^{\prime \prime }$ are symmetric, $%
A^{\prime }$ is self-symmetric under the exchange
\begin{equation}
(p_0,p_4)\Leftrightarrow (p_3,p_7).  \label{wee32}
\end{equation}
For $p_0+p_3\leq \frac 12,$ the parameter region is divided into four
subregions: $D_1^{\prime },D_1^{\prime \prime },D_2$ and $C_2.$ The regions $%
D_1^{\prime }$and $C_2$ are symmetric with each other in the sense
of the (\ref{wee32}). $D_1^{\prime \prime }$ and $D_2$ are the
biseparable subregions in the four parameter system. Notice that
when all of the inequalities (\ref{wee5})-(\ref{wee7}) are
preserved, we should consider the violation of inequalities
(\ref{wee8})-(\ref{wee10}). A similar analysis should be added for
the parameters $p_1,p_2,p_5,p_6$. The actual case is that
$D_1^{\prime \prime }$ and $D_2$ should be further divided into
truly biseparable subregions and genuine entangled subregions
according to the parameters $p_1,p_2,p_5,p_6$.

The genuine relative entropy of entanglement is analytically expressed as
five formulas, according to the subregion the genuine entangled state
belongs to. The five formulas can be further classified as three kinds: one
symmetric formula and two pairs. The pair formulas can be interchangeable
under the parameter exchange (\ref{wee32}).

\section*{Acknowledgement}

XYC and LZJ thank the funding by the National Natural Science Foundation of
China (Grant No. 60972071), Zhejiang Province Science and Technology Project
of China (Grant No. 2009C31060).

\section*{Appendix: Comparison of $E_B$ and $E_{A^{\prime \prime }}$}

Define a function
\[
E(x)=1-xH_2(\frac{p_3}x)-(1-x)H_2(\frac{p_0}{1-x}),
\]
then $E_B$ and $E_{A^{\prime \prime }}$ can be expressed as $E_B=E(x_B),$ $%
E_{A^{\prime \prime }}=E(x_{A^{\prime \prime }})$ with $x_B=p_3+p_7,$ $%
x_{A^{\prime \prime }}=1-p_0-p_4.$ Since $p_3+p_7+p_0+p_4\leq 1,$ thus
\[
x_B\leq x_{A^{\prime \prime }}.
\]
The derivative of the function $E(x)$ is
\[
\frac{dE(x)}{dx}=-\log _2\frac x{1-x}+\log _2\frac{x-p_3}{1-x-p_0},
\]
which leads to the solely extremal point
\[
x^{*}=\frac{p_3}{p_0+p_3}.
\]
At the extramal point $x^{*},$ the function $E(x)$ reaches its minimum $%
E(x^{*})$ since the second derivative at $x=x^{*}$ is positive. The function
$E(x)$ monotonically decreases with $x$ for $x\leq x^{*}$, $E(x)$
monotonically increases with $x$ for $x\geq x^{*}$.

Let the genuine entangled state be at regions $B,C$, we have $p_7>p_{AB}.$
We can rewrite it as $p_3+p_7>\frac{p_3}{p_0+p_3}.$ Hence $x^{*}<x_B\leq
x_{A^{\prime \prime }}.$ At the right side of $x^{*},$ the function $E(x)$
is a monotonically increasing function, so that
\[
E_B\leq E_{A^{\prime \prime }}.
\]
In region $B,$ the genuine entanglement is $E_B.$ In region $C,$ the genuine
entanglement is $E_C$ since $E_C\leq E_B\leq E_{A^{\prime \prime }}.$

Consider the regions $A^{\prime \prime },$ $A^{\prime \prime \prime }$ and $%
D_1^{\prime },$ we have $p_4>p_{A^{\prime }A^{\prime \prime }}.$ We rewrite
it as $p_0+p_4>\frac{p_0}{p_0+p_3},$ further we have $1-$ $p_0+-p_4<\frac{p_3%
}{p_0+p_3}$, which is $x_{A^{\prime \prime }}<x^{*}.$ Hence $x_B\leq
x_{A^{\prime \prime }}<x^{*}.$ At the left side of $x^{*},$ the function $%
E(x)$ is a monotonically decreasing function, so that
\[
E_B\geq E_{A^{\prime \prime }}.
\]
In regions $A^{\prime \prime }$ , the genuine entanglement is $E_{A^{\prime
\prime }}.$ In regions $A^{\prime \prime \prime }$ and $D_1^{\prime },$ the
genuine entanglement is $E_{A^{\prime \prime \prime }}$ since $E_{A^{\prime
\prime \prime }}\leq E_{A^{\prime \prime }}\leq E_B.$

\end{document}